# PMD-Tolerant 20 krad/s Endless Polarization and Phase Control for BB84-Based QKD with TDM Pilot Signals

Benjamin Koch[(1,2)], Reinhold Noé[(1,2)]

(1) Paderborn University, EIM-E, Warburger Str. 100, D-33098 Paderborn, Germany, e-mail: noe@upb.de
(2) Novoptel GmbH, Helmerner Weg 2, D-33100 Paderborn, Germany, E-mail: info@novoptel.com

*Abstract*—TDM-based polarization and differential phase control with 35ps PMD tolerance and 20krad/s tracking speed is demonstrated. 600ns intervals are reserved for QKD and for 0°- and 45°-polarized pilot signals. ECLs are modulated directly, with high extinction. Power budget is 17dB, fiber length is 63km.

## I. INTRODUCTION

Quantum key distribution (QKD) with the BB84 protocol [1, 2] based on polarizations requires polarization and differential phase control at the receive end. This means that not only an incoming polarization is to be mapped into a fixed polarization but also the phase difference between this polarization and its orthogonal must be controlled.

One such experiment is described in [3]. Two additional WDM pilot signals were used to stabilize the quantum channel. Significant degradation was observed at polarization scrambling rates >100 rad/s, and 400 rad/s was the highest tested value. In a similar setup, we have brought the polarization- and phase tracking speed to 20 krad/s [4, 5]. All result samples were good, not only their average. This proves that control was endless, i.e. uninterrupted. However, polarization mode dispersion (PMD) of the transmission fiber is a fierce enemy of polarization control with WDM pilot signals. TDM pilot signals are likewise possible [6]. No tracking speed was reported in [6] and there was no polarization scrambler or other external perturbation. A possibility to achieve TDM is to switch on cyclically a 0°, a 45° and a probe / quantum signal. This reduces achievable code rate by, say, a factor of 3 (assuming equal time slots for the 3 signals). The TDM scheme presented in the following has these advantages over WDM:

- Since the 3 optical frequencies can be chosen closer or equal, PMD tolerance is increased.
- TDM pilot signals from directly modulated lasers have high extinction. This improves channel quality.
- If TDM is added to WDM, filtering requirements and loss at the receive end are very much reduced, for increased reach or code rate.

## II. EXPERIMENTAL SETUP

Fig. 1 shows the experimental setup. An FPGA at the transmitter (TX) outputs 3 non-overlapping pulses. 3 identical commercial external cavity lasers (ECL) are modified to permit direct modulation. There is a 0°- and a 45°-polarized pilot laser, and a probe laser with variable polarization. The 3 lasers are gated or modulated with the 3 pulse signals. Outside their respective on-intervals the lasers are completely switched off, for best extinction during the probe interval. After being combined into one SMF the 3 signals are transmitted to the receiver. To test polarization fluctuation speed tolerance, there is a polarization scrambler 1 (Novoptel EPS1000) with rotating LiNbO3 electrooptic waveplates. To check also PMD tolerance, scrambler 2 and a differential group delay (DGD) section are added. DGD section and scrambler 2 can be replaced by 63 km of SMF.

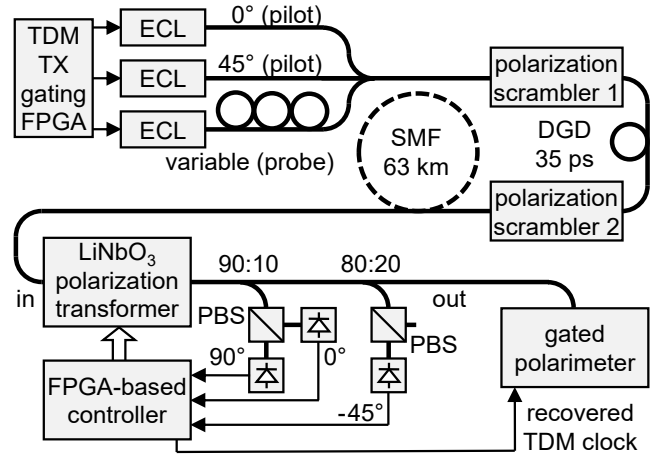

Fig. 1. Setup for PMD-tolerant polarization and phase control

In the receiver (RX) there is another LiNbO3 polarization transformer. Its electrooptic waveplates can endlessly transform an arbitrary polarization into a fixed one, and introduce endless differential phase shift between this polarization and its orthogonal. Endless means without interruption, no matter how long trajectories and how large phase shifts are. An FPGA serves as a controller. To obtain control signals there are two tap couplers, each with a polarization beamsplitter (PBS) at its output. At the 90° PBS output behind the 1st (10%) tap, a photodiode detects an error signal for conventional polarization control. Another photodiode at the 0° polarized PBS output behind the 1st tap detects the remaining power. The controller also adds these two photocurrents, thereby obtaining a replica of the total incident power for recovering of the TDM clock signal. Based on this info, the 90° signal is used for conventional polarization control only during the time interval in which the 0° pilot is switched on. The PBS behind the 2nd (20%) outputs a –45° polarized signal. After photodectection it serves as an error signal for differential phase control during the time interval in which the 45° pilot is switched on. Reason is the following: When the 0° pilot signal is transformed into 0° at the RX polarization transformer output the 45° pilot signal can be anything between 45°, right circular, –45°, left circular, depending on differential phase shift between 0° and 90° polarizations. Minimization of the –45° polarized error signal assures that the 45° pilot signal is received as 45°. The recovered TDM clock signal is output to a gated polarimeter (Novoptel PM1000). The polarization and phase controller is a modified version of a commercial subsystem/unit [7] that has made its way into the QKD field.

## III. RESULTS

In the TX, light is switched on for 600 ns in the 0° pilot, 600 ns in the 45° pilot and 560 ns in the probe laser. Period is 1800 ns. Clock recovery requires a low-light interval of at least 40 ns. Of course a low-power probe / quantum signal may stay on all the time because the pilot signals overshine it. The three lasers are tuned to (approximately) the same optical frequency 193.4 THz. Fig. 2 shows the sum intensity vs. time of the 3 periodically modulated lasers. This is measured with an extra tap coupler and an extra photodiode (not shown). The oscilloscope is triggered with the recovered TDM clock. The rear edge of one signal is also shown in 100-fold temporal close-up. Once the intensity has started to reduce, it reduces to zero in about 10 ns. One may expect a roughly exponential intensity decay at the very end of the optical pulse. Time constant looks to be <2 ns. This means that after several 10 ns an extreme extinction is achieved, better than achievevable with practical filters in a WDM setup.

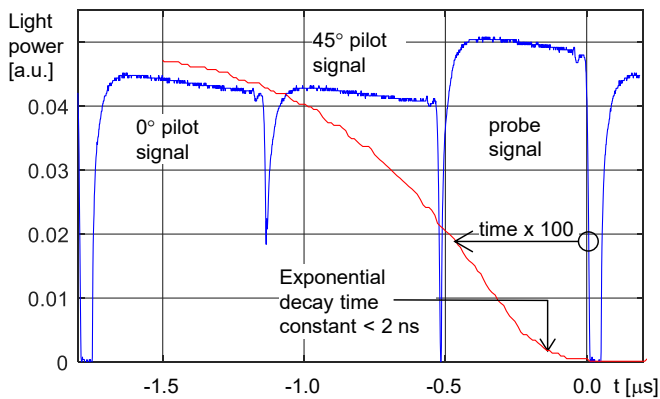

Fig. 2. Obtained light power waveform vs. time

For direct laser modulation, two schemes were tested: In one of them, the laser voltage was fixed at 0V during off-periods. In the other, the laser was reverse-biased at -0.7V during off-periods. Results were indistinguishable. This means that slower laser turn off effects seem not be present, since they would have been reduced when changing from 0V to -0.7V. To achieve a decent dynamic range, the 3 photodiodes are of avalanche type. Bias voltages are set in an initial automated procedure. Under this condition the allowed pulse peak power range of the pilots is −11.2 ... 0 dBm. With equal 0° and 45° pilot peak powers and weak probe power, mean input power of the polarization and phase controller can be −13 ... −1.8 dBm. Total insertion loss for the probe signal is ~5.5 dB, including a connector loss (not shown).

Next, control was switched on. The photocurrents detected in the 90° and −45° photodiodes were recorded during the time slots in which 0° and 45° polarizations were transmitted, respectively. Figs. 3-5 show complementary distribution functions 1−F(RIE) of relative intensity errors (RIE0°, RIE45°) indicated by these two photocurrents. The value on the logarithmic ordinate gives the probability that the relative intensity error surpasses the value of the abscissa. The distribution function of "no light" measurements serves to determine the zero point. The photocurrent behind the 90° PBS is minimized by two degrees-of-freedom (DOF) of the controller. An absolute minimum, RIE0° = 0, can be reached there. The third DOF is used to minimize the photocurrent behind the −45° PBS. A relative minimum RIE45° is achieved here, which can become RIE45° = 0 if the analyzed control polarizations (nominally 0° and 45°, blocked by the 90° and −45° PBS) span the same angle ($\pi/2$) on the Poincaré sphere as the transmitted control polarizations (nominally 0° and 45°).

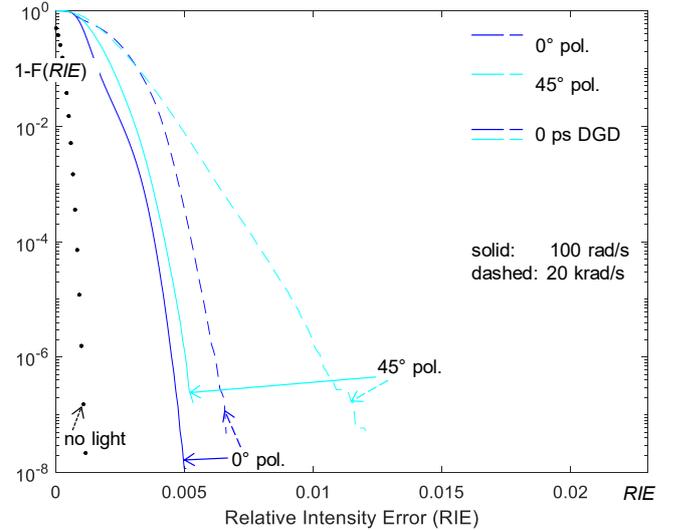

Fig. 3. Complementary distribution functions 1-F(RIE) of relative intensity errors (RIE) of 0° and 45° polarizations, measured at scrambling speeds of 0.1 and 20 krad/s over ~30 s, with 0 ps of DGD back-to-back. Ordinate value gives cumulative probability that a RIE surpasses the abscissa value.

Initially, only polarization scrambler 1 was in the link. Sets of measurements (RIE0°, RIE45°) were taken in about 30 s for polarization scrambling speeds of 100 rad/s and 20 krad/s. Results are given in Fig. 3.

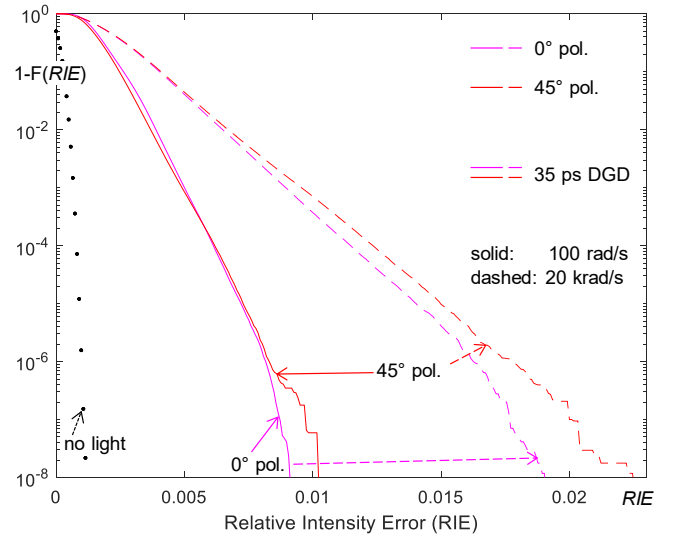

Fig. 4. Like Fig. 3, but with 35 ps of DGD back-to-back.

Then the 35 ps DGD element (made of PMF) and scrambler 2 were added and the experiment was repeated, see Fig. 4. In all cases, the smaller excursion of RIE0° vs. RIE45° indicate that transmitted 0°/90° are better controlled than ±45°/circular polarizations. Even for 35 ps of DGD at 20 krad/s scrambling, maximum RIE45° was fine for QKD purposes, up to 2.2%.

Time-resolved polarimetry behind the 35 ps DGD element showed a polarization change of 0.15 rad and hence a chirp of 700 MHz in each 600 ns long laser pulse, most of which occurred in the first pulse half. Since a probe / quantum signal with low power may stay on all the time and will hence have no chirp, PMD impact on it is expected to be lower than measured for our chirped pilot signals. We have assessed probe polarization stability with the gated polarimeter. 6 probe polarizations at the cartesian normalized Stokes axes were sequentially transmitted. Results are superimposed in Fig. 6, for 100 rad/s, 10 krad/s and 20 krad/s polarization scrambling (*a*, *b*, *c*), without PMD. In line with the observation $RIE45° > RIE0°$ in Figs. 3-5, the obtained spots are elliptical along the ±45°/circular great circle. For best performance of a BB84 QKD protocol, probe polarizations should be therefore 0°/90°/±45°, not ±45°/circular. The larger spot widths compared to [4, 5] are due to the 100 MS/s sampling rate of the polarimeter, as opposed to 1 kS/s sampling in [4, 5]. Conventional polarization control is also shown (*d*). As expected, only 0°/90° polarizations are controlled while ±45°/circular rotate on a great circle. With 35 ps of DGD, spots are broadened at 100 rad/s and 20 krad/s polarization scrambling (*e*, *f*). As mentioned, 0.15 rad of that broadening are due to the chirp of the probe pulse, which would be avoided in a QKD system.

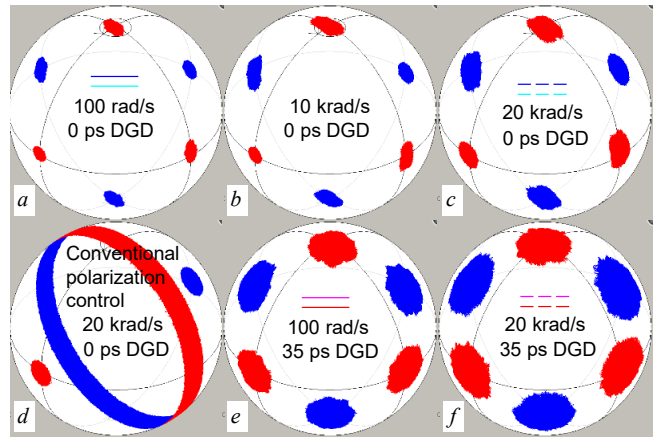

Fig. 6. Synopsis of 6 probe polarizations behind controller, accumulated over 5 s each, without (*a*, *b*, *c*) and with (*e*, *f*) PMD. Conventional polarization control is also shown (*d*).

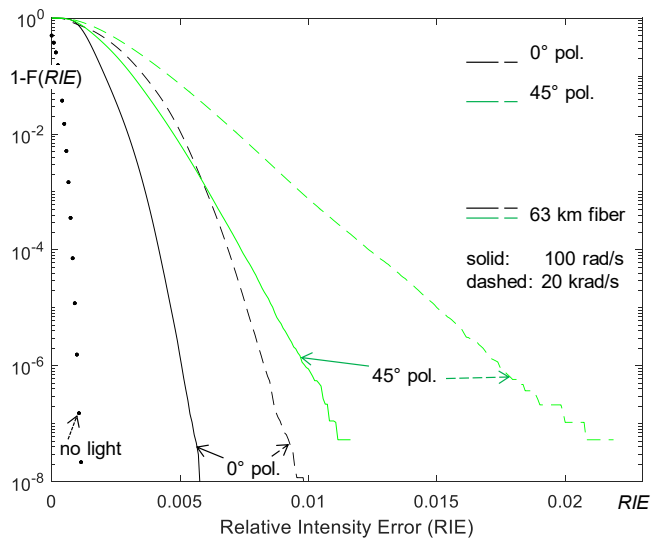

Fig. 5. Like Fig. 3, but with 63 km of SMF.

Finally, scrambler 2 and DGD element were removed and a 63 km long fiber with 14 dB loss was inserted. Probe signal and its coupler were likewise removed to maximize power budget, because practical quantum signals will have very low power anyway and can therefore be inserted via a coupler with very little through path loss. Link loss including polarization scrambler 1 is 17 dB. Received power is −13 dBm. Results are plotted in Fig. 5. Performance is similar to that obtained with 35 ps of DGD. 20 krad/s polarization scrambling is again permissible.

At least two possibilities exist to further improve extinction of pilot signals:

- Hybrid TDM/WDM approach in which the high TDM extinction relaxes the requirements on WDM filtering or permits closer WDM channel spacing.

- Much longer probe signal intervals, to let light scattering in fiber fully decay.

## IV. CONCLUSION

While polarization and phase control with WDM has made its way into the QKD field the present TDM-based implementation improves PMD tolerance and reduces insertion loss. We consider it as promising to implement large-scale polarization-based QKD networks for protection of sensitive information channels. The achieved polarization- and differential phase control speed of 20 krad/s seems to be orders of magnitude faster than previously obtained in QKD transmission with TDM pilot signals. Combination with the WDM approach is possible.